\documentclass[a4paper,fleqn]{cas-sc}
\usepackage{amsmath}
\usepackage{graphicx}
\usepackage{amssymb}
\usepackage{footnote}
\usepackage{amsfonts}
\usepackage{url}

\usepackage[numbers]{natbib}

\begin{document}

\let\WriteBookmarks\relax
\def\floatpagepagefraction{1}
\def\textpagefraction{.001}

\shorttitle{A quantitative framework for exploring exit strategies from the COVID-19 lockdown}
\shortauthors{A.S. Fokas, J. Cuevas-Maraver and P.G. Kevrekidis}

\title{A quantitative framework for exploring exit strategies from the COVID-19 lockdown}

\author[1,2]{A.S. Fokas}
\author[3,4]{J. Cuevas-Maraver}
\author[5,6]{P.G.\ Kevrekidis}

\address[1]{Department of Applied Mathematics and Theoretical Physics, University of Cambridge, Wilberforce Road, Cambridge CB3 0WA, U.K.}

\address[2]{Department of Civil and Environment Engineering, University of Southern
California, 90089, Los Angeles, Ca, USA}

\address[3]{Grupo de F\'{\i}sica No Lineal, Departamento de F\'{\i}sica Aplicada I,
Universidad de Sevilla. Escuela Polit\'{e}cnica Superior, C/ Virgen de \'{A}frica, 7, 41011-Sevilla, Spain}

\address[4]{Instituto de Matem\'{a}ticas de la Universidad de Sevilla (IMUS). Edificio Celestino Mutis. Avda. Reina Mercedes s/n, 41012-Sevilla, Spain}

\address[5]{Department of Mathematics and Statistics, University of Massachusetts, Amherst, MA 01003-4515, USA}

\address[6]{Mathematical Institute, University of Oxford, Oxford, UK}

\begin{abstract}
Following the highly restrictive measures adopted by many countries for combating the current pandemic, the number of individuals infected by SARS-CoV-2 and the associated number of deaths  steadily decreased. This fact, together with the impossibility of maintaining the lockdown indefinitely, raises the crucial question of whether it is possible to design an exit strategy based on quantitative analysis.  Guided by rigorous mathematical results, we show that this is indeed possible: we present a robust numerical algorithm which can compute the cumulative number of deaths that will occur as a result of increasing the number of contacts by a given multiple, using as input only the most reliable of all data available during the lockdown, namely the cumulative number of deaths.
\end{abstract}

\begin{keywords}
COVID-19 Modeling \sep Ordinary Differential Equations \sep Cumulative Death Modeling \sep Parameter Estimation and Optimization \sep Lockdown Exit Strategies
\end{keywords}

\maketitle

\section{Introduction}

On December 31, 2019, the Chinese government reported a cluster of pneumonia cases of unknown cause that was later identified as a result of the severe acute respiratory syndrome coronavirus 2 (SARS-CoV-2). This is one of the most serious manifestations of an infection and associated disease (termed COVID-19) caused by a coronavirus.
Like earlier outbreaks caused by two other pathogenic human respiratory coronaviruses, namely, the severe respiratory syndrome coronavirus (SARS-CoV) and the Middle East respiratory syndrome coronavirus (MERS-CoV), SARS-CoV-2 causes a respiratory disease that it is often severe. In addition, SARS-CoV-2 can attack many vital organs of the body, and can also lead to severe neurological disorders, including the Guillain-Barr{\'e} syndrome~\cite{dal2,dalakas}. Fortunately, SARS-CoV-2 is associated with lower mortality than its above predecessors; however, it is more contagious~\cite{Munster}. As a result of this fact and the lack of early measures for curtailing its spread, it  has caused a pandemic, which is considered as the most serious threat to public health since the pandemic caused by the  1918 influenza (the 1918 pandemic, which is  the  deadliest event in human history, caused more than 50 million deaths which corresponds to 200 million deaths in today’s population).

The scientific community is playing a crucial role in combating the above threat: from elucidating fundamental features of SARS-CoV-2 and mechanisms of its transmissibility~\cite{elife}, to addressing the vital question of a pharmacological treatment and the development of an effective vaccine~\cite{oxford}. For example, the viral genome of SARS-CoV-2 has been sequenced~\cite{Zhu}. Also, mechanisms underlying   the increased transmissivity of SARS-CoV-2 have been traced to its dual receptor attachment in the host cells. In particular, it has been shown that the attachment of
the virus to the surface of respiratory cells is mediated by certain viral proteins which bind not
only to the angiotensin converting enzyme-2 (ACE-2) receptor~\cite{FCM1}, but also to sialic acid containing glycoproteins and gangliosides that reside on cell surfaces~\cite{FCM2}~\footnote{This is to be contrasted with the case of SARS-CoV, which binds only to ACE-2 receptors~\cite{FCM1,FCM2}.}. Regarding pharmacological interventions, a randomized, controlled study involving hospitalized adult patients with confirmed SARS-CoV-2 infection, showed that there was no benefit from the antiviral regime of  lopinavir-ritovir (which is an effective treatment in patients infected with human immunodeficient virus)~\cite{Cao}. Similarly, the combination of the anti-malarial medication chloroquine, and its derivative hydroxychloroquine, with or without the antibiotic azithromycin, may not only be ineffective, but also potentially harmful~\cite{Magagnoli,CQ_lung,HCQ_macaques} (see also the meta-analysis \cite{HCQ_meta}).

In addition, the scientific community has had a vital impact in decisions made by policy makers of possible non-pharmacological
approaches
to limit the catastrophic impact of the pandemic. For example, two possible strategies, called mitigation and suppression, are thoroughly discussed in~\cite{lev3, newer}; in the early stages of the pandemic, UK was following mitigation, but after the publication of this report, it adopted suppression. This policy, which was already implemented in several other countries (of course there were also notable exceptions such as Sweden
and Brazil with serious associated consequences), has led to the curtailing of the pandemic. Indeed, the number of individuals reported to be infected with SARS-CoV-2, as well as the number of related deaths, steadily decreased in the countries that  adopted severely restrictive measures, known as ‘lockdown’; see e.g.
the data in~\cite{jhu}. This welcome development, together with the impossibility of maintaining indefinitely the lockdown conditions (both for obvious economical reasons and for the induced serious psychological effects) raises the following question: is it possible to design an exit strategy based on the quantitative analysis of the effect of easing the lockdown measures? The purpose of this work is to provide an affirmative answer to this question, concentrating on the number of associated deaths, and
the unique identification for a canonical epidemiological model of parametric combinations
that determine it.

We note in passing that the use of mathematical methods in a plethora of
biological problems at the level of (both discrete and continuum)
model analysis and at that of finding a wide variety
of special solutions is a theme of intense interest in recent years. As some relevant
examples, we mention the works of~\cite{osm1,osm2} (while similar methods
have been used in nonlinear engineering and mathematical physics
problems~\cite{osm3,osm4,osm5,osm6}). In the recently very highly active
front of COVID-19 modeling, some of the efforts have been directed
at modeling the early stages of the pandemic~\cite{kax1};  others have focused
on designing a pandemic response index (to quantify/rank the response
of different countries)~\cite{kax2} or towards quantifying the response
of different regions within a country, e.g., the states within the USA~\cite{kax3}.
Similarly, models have focused on cruise ships~\cite{mizumoto}, on
cities~\cite{beijing},
as well as states/provinces~\cite{beijing,england,louisiana,arenas},  but also various
countries~\cite{newer,sweden,tsironis,albania, brazil,marma}, aside
from the prototypical examples of Wuhan, China~\cite{mbe},
and some among the hard-hit Italian provinces~\cite{siettos}.

Before providing details of our proposed methodology for the  computation of the number of deaths following a specific increase of contacts between asymptomatic individuals infected with COVID-19, it should be emphasized that the answer to the above question is literally a matter of life and death: (i) No therapeutic intervention has been proven so far effective for the treatment of the severe illnesses and side-effects caused by SARS-CoV-2. (ii) Reported mortality rates differ drastically between different countries and are crucially affected by age. For example, in the largest study from China involving 1099 hospitalized patients with laboratory confirmed SARS-CoV-2
infection of median age 47 years, only 5\% needed admission to the intensive care unit, 2.3\% underwent invasive mechanical ventilation, and 1.4\% died~\cite{Guan}. On the other hand, following the identification on February 28, 2020 of a confirmed case of COVID-19 in a nursing facility in Washington, USA, as of March 18, 101 residents of this facility and 50 health care personnel were confirmed with COVID-19, and in addition 16 infected visitors  were epidemiologically linked with this facility. Hospitalization rates for residents, staff, and visitors, were 54.5\%, 6\%, and 50\% respectively; the corresponding mortality rates were 33.7\%, 0\%, and 6.2\%~\cite{Michel}. (iii) Despite the fact that the unprecedented efforts for the development of a vaccine take place within a framework of explosive progress in basic scientific understanding that has occurred in the areas of genomics and structural biology, it is not expected that a vaccine will be available for at least several
months~\cite{lurie}. From the above remarks it becomes clear  that the lockdown measures must be eased without the benefit of any substantial pharmacological cover, which is desperately needed especially for older persons (and individuals  with a variety of diseases such as Diabetes type 2, hypertension, and respiratory disorders), and without the anticipation of the imminent availability  of a vaccine.

There is no doubt that the solution to the vitally important problem of ‘how to defeat the current pandemic’, will finally be provided by medicine and biology via the development of appropriate pharmacological treatments and an effective vaccine. However, it appears that at the moment, there exists a unique opportunity for mathematical modeling and associated analysis to contribute definitively to the partial
addressing of this problem. In particular, SIR (susceptible, infected, recovered) type models are widely accepted in mathematical epidemiology~\cite{SIR}; and  this type of models can be modified to capture some key features of
the present pandemic (such as the crucial role of asymptomatically
infected individuals~\cite{mizumoto, arons2020,asympt}).
Thus, it follows that the rigorous analysis of such a class of models provides a possible
approach towards studying quantitatively the effect of easing the lockdown
measures. If such analysis can be leveraged to give rise to accurate and reproducible computational results based on reliable data, then it would be possible to
explore systematically the design of a safe lockdown exit.

The presentation of our efforts in the above direction is structured as follows. In section 2, we provide an overview
of the computational algorithm that we propose, as well as its epidemiological
implications towards identifying the model parameters and also evolving the
model in the future towards release of lockdown conditions. Then, in section
3, we delve into the details of our mathematical methods and corresponding
computational results. Finally, in section 4, we summarize our
findings and discuss a number of future recommendations and interesting
research directions.

\section{A computational algorithm based on a rigorous mathematical result}

The model adopted in this work, which is discussed in detail in the Methods and
Results section, involves 6 (first order) ordinary differential equations (ODEs) uniquely specified by 9 constant parameters (and, of course, well-posed under the introduction of $6$ associated
initial conditions). One of the relevant parameters, namely the constant $c_1$, denotes the effect of the interaction of the asymptomatic, infected individuals with
those susceptible to be infected. This parameter is particularly important for our analysis: the easing of the lockdown conditions would result in increasing the number of contacts (to which $c_1$ is proportional) among asymptomatic and
susceptible individuals, and this effect can be straightforwardly incorporated
in the model by replacing $c_1$ with $\zeta c_1$, where $\zeta$ is a fixed number, such as $2$ or $3$; this will be referred to, respectively, as the doubling or tripling of the number of contacts.

How can our model yield a computational approach to designing a safe lockdown exit? Suppose that, somehow, given appropriate data obtained during the lockdown period, the values of the $9$ parameters specifying the model could be determined
(and the associated initial conditions prescribed). Then, using these values, replacing $c_1$ by $\zeta c_1$, and solving the resulting 6 ODEs, the number of infected, hospitalized, recovered, and deceased individuals in the post-lockdown period could be computed. It turns out that the above scenario, which would yield information about all the basic features of the post-lockdown state, is impossible.  An implicit assumption in this class of epidemiological models is that the model parameters can be uniquely determined from an appropriate set of data. However, this is far from a
trivial assumption and the whole branch of identifiability of the models
is associated with this issue~\cite{eisenberg}. In this connection it should be noted that the problem of determining the model-parameters from  the given data can be formulated as an inverse problem, and such problems are notoriously difficult, especially regarding the important question of uniqueness; namely, proving that the given data give rise to a unique set of parameters\footnote{For example, it is rigorously established in~\cite{fokas} that different neuronal electric currents give rise to the same data obtained via electroencephalographic (EEG) recordings. Thus, it is impossible to obtain uniquely the current via the solution of the inverse problem associated with EEG.}. Actually, we have shown that it is impossible to determine uniquely all 9 model parameters from a reliable set of given data.
This is indeed an example of dimension reduction: this represents the crucial aspect
of whether a given model outcome depends not on individual parameters
 alone (except for one of the model parameters) but rather on combinations formed by suitable (irreducible) combinations of
 parameters; for a recent, data-driven example, see~\cite{yannis}.
In addition to the existence of the above prohibitive result, many of the needed data are unavailable. For example, although the model involves the total number of infected individuals, the available data are for the ‘reported infected’ individuals.

By concentrating on the number of deaths, we have been able to overcome both of the above difficulties. Indeed, we have shown that it is possible to determine the single model parameter and the 6 specific combinations of the 9 model parameters that characterize a certain 4th order ODE specifying the time-evolution of the number of deaths. Furthermore, we have
established that this can be achieved uniquely by using the most reliable of all available data, namely, the data for the cumulative number of deaths.

The above mathematical results give rise to the following algorithm (which we have
showcased in a number of select examples):  starting with the death data during the lockdown period, we compute uniquely the single model parameter as well as the 6 combinations of the original 9 model parameters that specify the 4th order ODE determining the evolution of the number of deaths. The parameter $c_1$ enters one of these combinations, and fortunately, it enters in a homogeneous manner; thus, replacing $c_1$ by $\zeta c_1$ results in replacing the relevant combination by $\zeta$ times its original value. Then, using this combination, the single model parameter, and the remaining 5 combinations determined from the death data, we can proceed to forward time-step the resulting   4th order ODE. This uniquely yields the number of deaths in the post-lockdown period (the concrete procedure used to test the robustness of our algorithm, is described in the Methods and Results section).

We applied the above approach to the epidemics of the countries of
Portugal and Greece, as well as the autonomous community of Andalusia in
Spain. These have been selected due to their notable similarities
(geographic location, as well as similar populations in the vicinity of
8-10 million residents), but also due to their significant differences
in connection to the response to the pandemic. Greece has had an extremely
low tally of deceased individuals as a result of the pandemic,
being one of the notable success
examples of early application of lockdown measures~\cite{time}.
Portugal is also a case with relatively low cumulative numbers,
although, at around the same population as Greece, it currently has
more than 7 times the number of deaths. Andalusia, on the other hand,
has a smaller population (by about 2 million) than the other 2, but has
already $1391$ reported deaths (i.e., $49$ more than Portugal
and more than 8 times more than Greece). Hence, these are interestingly
distinct examples.
In the case of Greece, and through the findings reflecting the
low transmissivity of the virus, we find that a doubling of the
contacts in the entire population, i.e., $\zeta=2$, would only slightly
increase the number of deaths from 158 to 167. On the other hand,
the effect of such a widespread easing of the lockdown measures would be quite
different in the case of Portugal (from 1433 to 40727) and of Andalusia (from 1451 to 26846).
Furthermore, far more dramatic is the impact in the case of changing
$\zeta$ to $3$. In that case, the number of deaths is drastically
increased even in Greece (18474); the situations in Portugal (84014) and Andalusia (106855) may become catastrophic.
The derivation of the above results is  the consequence  of a stable numerical algorithm capturing  the   key finding presented here, namely,  that it is possible  to: (i) identify the irreducible sets of parameters associated with the system; and (ii) to subsequently utilize
(only) those combinations towards the  forward prediction of the death tally upon
different scenarios of easing the  lockdown measures.

\section{Methods and Results}

\begin{figure}[!ht]
\begin{center}
   \includegraphics[width=.45\textwidth]{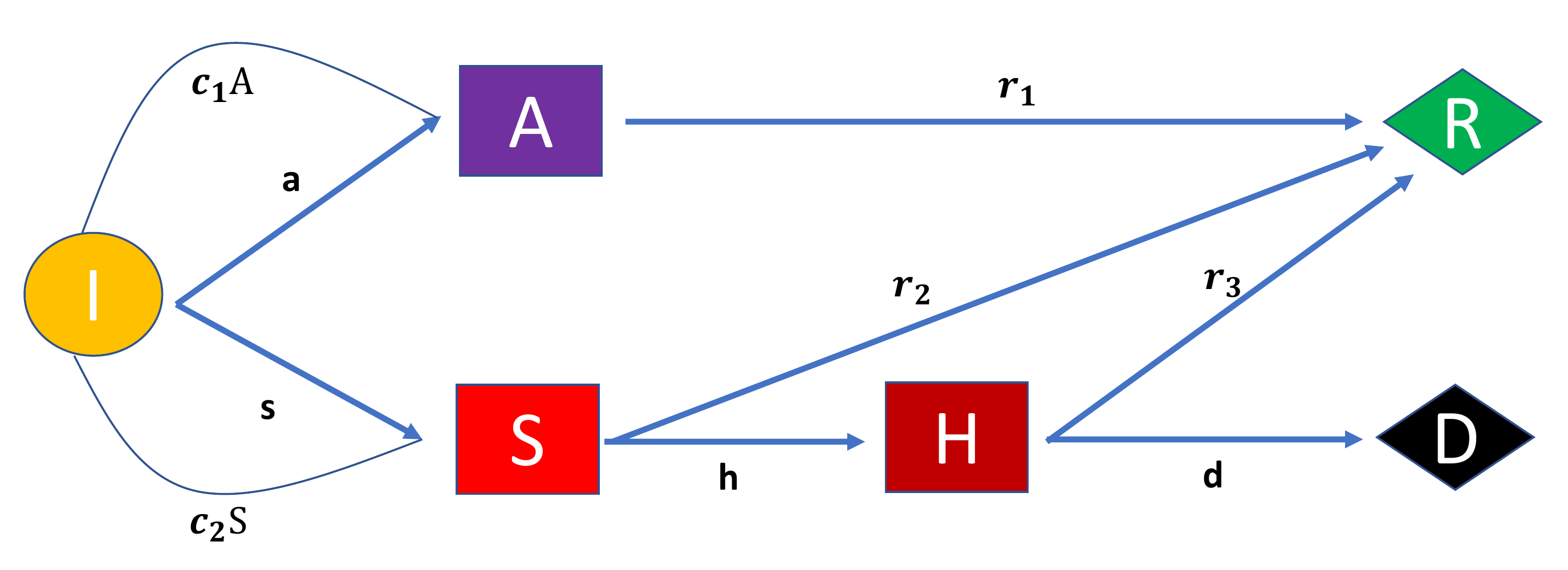}  \\
\end{center}
\caption{Flowchart of the populations considered in the model and the rates of transformation between them. The corresponding dynamical equations are
  Eqs.~(\ref{eqn1})--(\ref{eqn0}).}
\label{un_f0}
\end{figure}

Let $I(t)$ denote the infected (but not infectious) population. An individual in this population, after a median 5-day period (required for incubation~\cite{arma}) will either become sick or will be asymptomatic, as shown in the
flowchart  of Fig.~\ref{un_f0}~\footnote{An interval of
  3-10 days captures 98\% of the cases.}, which represents all the relevant populations
and the transitions between them. The sick and asymptomatic populations will be denoted, respectively, by $S(t)$ and $A(t)$. The rate at which an infected person becomes asymptomatic is denoted by $a$; this means that each day $a I(t)$ persons leave the infected population and enter the asymptomatic population. Similarly, each day $s I(t)$ leave the infected population and enter the sick population.
The asymptomatic individuals recover with a rate $r_1$, (i.e., similarly each day  $r_1A(t)$ leave the asymptomatic population and enter the recovered population), which is  denoted by $R(t)$. The sick individuals either  recover with a rate $r_2$ or they become hospitalized $H(t)$ with a rate $h$. In turn, the hospitalized patients also have two possible outcomes of the medical intervention efforts; either they recover with a rate $r_3$, or they become deceased, $D(t)$, with a rate $d$.

It is straightforward to write the above statements in a mathematical form; this gives rise to the equations below:

\begin{eqnarray}
\frac{dA}{dt} &=& a I - r_1 A,  \label{eqn1} \\
\frac{dS}{dt} &=& s I - (h + r_2) S, \label{eqn2} \\
\frac{dH}{dt} &=& h S - (r_3+d) H, \label{eqn3} \\
\frac{dR}{dt} &=& r_1 A + r_2 S + r_3 H, \label{eqn4} \\
\frac{dD}{dt} &=& d H. \label{eqn5}
\end{eqnarray}

In order to complete these equations, it is necessary to describe the mechanism via which a person can become infected. For this purpose, we follow the standard assumptions made in the SIR-type~\cite{SIR} epidemiological models: let $T$ denote the total population and let $c$ be the transmission rate proportional to the number of contacts per day made by an individual with the capacity to infect. Such a person belongs to $S$, or $A$, or $H$. However, for simplicity we assume that the hospitalized population cannot infect; this assumption is based on two considerations: first, the strict protective measures taken in the hospital, and second, the fact that hospitalized patients are infectious only for part of their stay in the hospital. The asymptomatic individuals are (more) free to interact with others, whereas the (self-isolating) sick persons are not; yet, the viral loads of the two have been argued to be
similar~\cite{arons2020}. Thus, we use $c_1$ and $c_2$ to denote the corresponding transmission rates (per person in the respective pools) of the asymptomatic and sick respectively. The number of persons available to be infected is $T-(I+S+A+H+R+D)$. Indeed, the susceptible individuals consist of the total population minus all the individuals that after going through the course of some phase of infection, either bear the infection at present ($I+A+S+H$) or have died from COVID-19 ($D$) or are assumed to have developed (even if temporary, but sufficient for our time scales of interest) immunity to COVID-19  due to recovery ($R$). The rate by which each day individuals enter $I$ is given by the product of the above expression with $c_1A+c_2S$. At the same time, as discussed earlier, every day $(a+s)I$  persons leave the infected population. Thus, the rate of change of $I(t)$ reads:
\begin{eqnarray}
  \frac{dI}{dt} = \left[T - (I+S+A+H+R+D)\right] \left(c_1 A + c_2 S\right) - (a+s) I. \label{eqn0}
 \end{eqnarray}
The above model depends on the given (total population) constant $T$ and on the 9 parameters $d$, $h$, $s$, $a$, $c_1$, $c_2$, $r_1$, $r_2$, $r_3$.

The {\it fundamental} inverse problem that we consider herein is the following: which specific parameters and combinations of the model parameters can be uniquely determined from the knowledge of the death data? We utilize here the data on the deceased individuals because it is the most reliable time series among the ones available. Indeed, there exists a significant uncertainty regarding the number of true infections (vs. the ones officially reported). This uncertainty is even more significant with respect to the current situation regarding the asymptomatics which, while concretely studied in some cases~\cite{mizumoto}, presents tremendous variability~\cite{asympt} between studies. As indicated above, the issue at hand is crucial from the point of view of modeling, both regarding issues of dimension reduction~\cite{yannis}, as well as those of identifiability of the models~\cite{eisenberg}.

It turns out that the cumulative number of deceased, $D$, satisfies a 4th order nonlinear ODE, which is uniquely determined by  the constants $\alpha= hsd(T- \mu)$ and  $\beta$, where $\mu$ and $\beta$ are integration constants, the model parameter $r_1$, as well as  the following 5 combinations of the model-parameters:  $C_1=ac_1/(h s d)$, $C_2=c_2/(h d)$,  $F=a+s$,
$R_2=r_2+h$,  $R_3=r_3+d$.
It is relevant to note here, in passing, that the initial condition dependent $\mu \ll T$ practically, and hence it is possible through $\alpha$, in principle, to provide a good approximation
of the product $h s d$.
This  basic ODE assumes the form
 \begin{eqnarray}
   q_4+ q_3 \left(q_2 + r_1 {\rm ln}(|q_3|) \right)=0,
   \label{eqn7}
   \end{eqnarray}
where
\begin{eqnarray}
  q_2 &=& (C_1+C_2) D^{(2)} + \left[C_1 (R_3+R_2) +C_2 (R_3+r_1)\right] D^{(1)}
  + R_3 (C_1 R_2 + r_1 C_2) D + \beta,
  \nonumber
  \\
  q_3 &=& D^{(3)} + k_3 D^{(2)} + k_2 D^{(1)} + k_1 D -\alpha,
  \nonumber
  \\
  q_4 &=& D^{(4)} + k_3 D^{(3)} +k_2 D^{(2)} + k_1 D^{(1)},
  \label{eqn8}
\end{eqnarray}
with superscripts denoting the number of derivatives with respect to $t$, and the constants appearing in Eq.~(\ref{eqn8}) defined as follows:
\begin{eqnarray}
  C_1 &=& \frac{a c_1}{h s d}, \quad C_2=\frac{c_2}{h d}, \quad R_2=r_2 + h,
  \quad R_3=r_3 + d, \\
  \nonumber
  k_1&=&F R_2 R_3, \quad k_2=F (R_2+R_3)+ R_2 R_3, \quad
  k_3= F + R_2 + R_3, \quad F=a+s.
  \label{eqn9}
\end{eqnarray}

We next explain how the central equation~(\ref{eqn7})
has been obtained and what the practical implications of the above  results
are before  turning to the examination of our numerical findings.
The main idea is to work backwards and utilize Eq.~(\ref{eqn5})
as a way to express the $H(t)$ in terms of $D^{(1)}$, then
Eq.~(\ref{eqn3}) to express $S(t)$ in terms of $D^{(1)}$ and
$D^{(2)}$, then Eq.~(\ref{eqn2}) to express $I(t)$ in terms of
$D^{(1)}, D^{(2)}$ and $D^{(3)}$; do the same for $A(t)+R(t)$ etc.
Reaching back at the level of Eq.~(\ref{eqn0}), one possible direction is to solve this
equation as an algebraic one for $A$ and then substitute back
to Eq.~(\ref{eqn1}). This leads to an interesting in its own right nonlinear 5th order
ODE. However, one can obtain an even more substantial and quite
remarkable simplification, as a result of the structure of the system.
It turns out that the quantity $[T - (I+S+A+H+R+D)]$ can be
expressed in terms of the $D(t)$ time series according to the
form provided by $q_3$ above; it is a direct calculation to see
that $dI/dt + (a+s) I$ is essentially the derivative of this expression
that enters into $q_4$. As a result, the ratio of these expressions $q_4/q_3$
becomes a logarithmic derivative, enabling us, when utilized together
with Eq.~(\ref{eqn1}), to identify a
‘first integral’ of the relevant equation that ultimately results into
Eq.~(\ref{eqn7}) at the mere expense of introducing an additional integration
constant ($\beta$). In fact, in addition to the remarkable simplification of
obtaining this as a 4th order ODE system, the presence of $\beta$ offers
a useful benchmark for the numerical method as we will see below.

There is an additional, important mathematical observation that concerns
the stability calculation of the healthy state of the system which in our
model is represented by $(I,S,A,H,R,D)=(0,0,0,0,0,0)$ i.e., the null state.
The spectral stability analysis of this state determines whether an epidemic
will grow or decay on the basis of the dominant stability eigenvalue $\lambda$ of
the relevant state. The destabilization threshold $\lambda=0$ corresponds
to a basic reproduction number $R_0=1$~\cite{SIR}. Positive $\lambda$ leads to
$R_0>1$ and spreading of the epidemic, whereas $\lambda<0$ leads to
the epidemic subsiding over time ($R_0<1$). Starting from a healthy state
in which case $\mu=0$ and $\alpha=hsd T$ in the expressions above, we find that the
relevant eigenvalue can be obtained by the largest eigenvalue of the polynomial:
\begin{eqnarray}
  \lambda^3+ \lambda^2 (F+r_1+R_2) +\lambda \left[F (r_1+R_2)+r_1 R_2 -\alpha (C_1 + C_2)\right]
  + F r_1 R_2 - \alpha (C_1 R_2 + C_2 r_1)=0.
  \label{evalue}
\end{eqnarray}
Remarkably, the relevant dominant eigenvalue depends {\it solely}
on the reduced quantities (including, of course, the role of
the total population introduced through
$\alpha$) of the original system.
In fact, the direct application of the Routh-Hurwitz conditions~\cite{routh}
leads to the criterion
\begin{eqnarray}
  \alpha (C_1 R_2 + C_2 r_1) <  F r_1 R_2,
  \label{evalue2}
\end{eqnarray}
for the subsiding of the epidemic (while the opposite sign of the inequality will
lead to its spreading, and equality will lead to a vanishing eigenvalue or
$R_0=1$).
The above result shows that $r_1$ and the reduced set of parametric combinations identified earlier, not only determine the most important tally of the impact of the epidemic,
namely the number of deaths, but also the potential capacity of the epidemic to spread.

It is important to mention that a similar analysis regarding the evolution of $D(t)$   can be performed in the case of two subpopulations, young and older~\cite{FCK}. Then,
Eq.~(\ref{eqn7}) is replaced by the two equations
\begin{eqnarray}
  q_4^l + P q_3^l=0,
  \label{eqn10}
  \end{eqnarray}
  where $q_3^l$ and $q_4^l$, are defined via the equations obtained from
  (\ref{eqn8}) by simply inserting superscripts y or o, whereas P is defined by:
  \begin{eqnarray}
    P=\sum_{l=y,o} C_1^l A^l + C_2^l S^l.
    \label{eqn11}
  \end{eqnarray}
  Here, $A^l$ and $S^l$ correspond to the asymptomatic and sick populations
  of each pool (young or older) that can be algebraically expressed
  in terms of $D(t)$ of the two sub-populations and their derivatives.

    Our theoretical analysis has provided the lumped (reduced) sets
    of coefficients in the spirit of dimension reduction that can be {\it uniquely}
    identified on the basis of the time series of $D(t)$, namely
    $r_1$, $C_1=ac_1/(h s d)$, $C_2=c_2/(h d)$,  $F=a+s$,
    $R_2=r_2+h$,  $R_3=r_3+d$. Additionally, one can identify the
    two integration constants $\alpha$ and $\beta$. In the practical
    implementation of our algorithm we will utilize the knowledge
    of the existence of such an ``optimal'' set of (lumped) parameters
    to leverage the ODE of Eq.~(\ref{eqn7}) to identify these parameters.
    The idea is that given a prescribed time-series for $D(t)$ and its
    derivatives, the left- hand side of Eq.~(\ref{eqn7}) can be evaluated.
    For a perfect time-series and an exact integration procedure,
    the unique global optimum parameter set would render the
    time series of this left-hand side (that we will refer to as $f$ hereafter)
    vanishing. As a measure of its vanishing, since practically the
    time series amounts to a vector at given times, we will use
    the $||f|| \equiv ||f||_{l^2}$; other norms can be used as well.

    Despite the existence of  the above well-defined mathematical procedure, there arise
    multiple practical complications. In particular,
    even if one has a ‘perfect’ time series
there is still the approximate nature of numerical integration. In what follows, we will generate such  a perfect series, that we will refer
    to as ``synthetic'' hereafter, via the integration
    of Eqs.~(\ref{eqn1})-(\ref{eqn0}) for given (prescribed) parameters.
    In practice, computations have a finite numerical accuracy that will be reflected
    in the approximate as opposed to the  exact vanishing of $||f||$. Naturally,
 the situation with practical data-based time series, is far more
    difficult. In that case, we will only know $D$ and we need to infer
    its (noisy) first four time-derivatives, before substituting all the relevant time
    series ($D, D^{(1)}, D^{(2)}, D^{(3)}$ and $D^{(4)}$) into $f$, in order  to
    perform the minimization.

We explored two possibilities in order
    to examine whether our approach could  be brought to bear in
    practical situations. In the first one, we considered finite
    difference-based approximations of the prescribed time series for $D(t)$;
    $D(t)$ constitutes the only provided reliable
    data to be used in the inverse-problem formulation for the identification of the relevant
    parameters. In the second approach we
focused on a logistic formula approximation earlier explored systematically,
    e.g., in~\cite{FDK1,FDK2}. In the latter case, we can consider
    the variable $x=D/D_F$ (where $D_F$ is the (asymptotically) final  number of
    deceased individuals); then , the differential equation (7) becomes
    \begin{eqnarray}
      x^{(1)}=kx(1-x) \Rightarrow x^{(2)}=k^2 (x-3 x^2 +2 x^3) \Rightarrow x^{(3)}=
      k^3 (x- 7 x^2 + 12 x^3 - 6 x^4) \Rightarrow x^{(4)}=\dots
      \label{eqn12}
      \end{eqnarray}
      The advantage of the second  methodology is its simplicity when employed
      in practical data-driven settings, as well as the smoothness of the quantities involved.
      Namely, the given time series for $D(t)$ is fitted into a sigmoid form:
      $D(t)=D_F /(1 + b e^{-k t})$, where $b$ (for given $D_F$) is essentially
      related to the initial condition $D(0)=D_F/(1+b)$ and $k$ is the truly
      essential (and most robust~\cite{FDK1,FDK2}) piece of information
      of the sigmoid fit that
      enters the differential equations of~(\ref{eqn12}). Substitution of
      these ODEs into Eq.~(\ref{eqn7}) ultimately leads to a time series $f$,
      or upon expansion (using  the dominance of the term involving $T$
      within the logarithmic term) to a polynomial. The minimization of
      this time series (in what is shown below) or of the associated polynomial
      was used for identifying the parameters by employing a procedure to be explained below.
      Naturally, as is the case with any such approximate method, there are also limitations
      to be discussed in our numerical examples that follow.

\begin{table}
\centering
\begin{tabular}{cccc}
Parameter & Exact & Finite differences & Sigmoid\\
\hline
$C_1$ & $2.3664\times10^{-5}$ & $2.3663\times10^{-5}$ & $2.3667\times10^{-5}$\\
$C_2$ & $5.0891\times10^{-6}$ & $5.0890\times10^{-6}$ & $5.0761\times10^{-6}$\\
$R_2$ & $0.2451$ & $0.2451$ & $0.2451$\\
$R_3$ & $0.1309$ & $0.1309$ & $0.1309$\\
$F$ & $0.3815$ & $0.3815$ & $0.3815$\\
$\alpha$ & $1387.7$ & $1387.7$ & $1392.1$\\
$r_1$ & $0.1482$ & $0.1482$ & $0.1482$ \\
$\beta$ & $-1.0719$ & $-1.0719$ & $-1.0723$\\
\hline\\
\end{tabular}
\caption{{Lumped parameter combinations (first column) of the local, but also global
    within our considerations, minima for the synthetic time
    series example and their numerical values for the exact up to numerical accuracy case (second column),
  the  finite difference derivative approximation (third column) and the sigmoid
  approximation (fourth column) approaches.}}
\label{tab1}
\end{table}

\begin{table}
\centering
\begin{tabular}{ccc}
$\overline{C_1}$ & $\overline{C_2}$ & $||f||$ \\
\hline
$1$   & $1$   & $5.47\times10^{-9}$ \\
$1/1.1$   & $1.4$ & $3.13\times10^{-5}$ \\
$1.1$ & $1/1.9$ & $3.04\times10^{-5}$ \\
$1/1.2$ & $1.2$ & $4.16\times10^{-4}$  \\
$1.2$ & $1/1.6$ & $4.79\times10^{-4}$  \\
\hline\\
\end{tabular}
\caption{For the exact (up to numerical accuracy) case, the value of the global
  optimum and of the ``next best things'' (first two columns) and its corresponding fitness (third column) are
  given. The first two columns are given in terms of multiples of the $(C_1,C_2)$
pair provided in Table 1.}
\label{tab2}
\end{table}

\begin{figure}[!ht]
\begin{center}
\begin{tabular}{ccc}
\includegraphics[width=.3\textwidth]{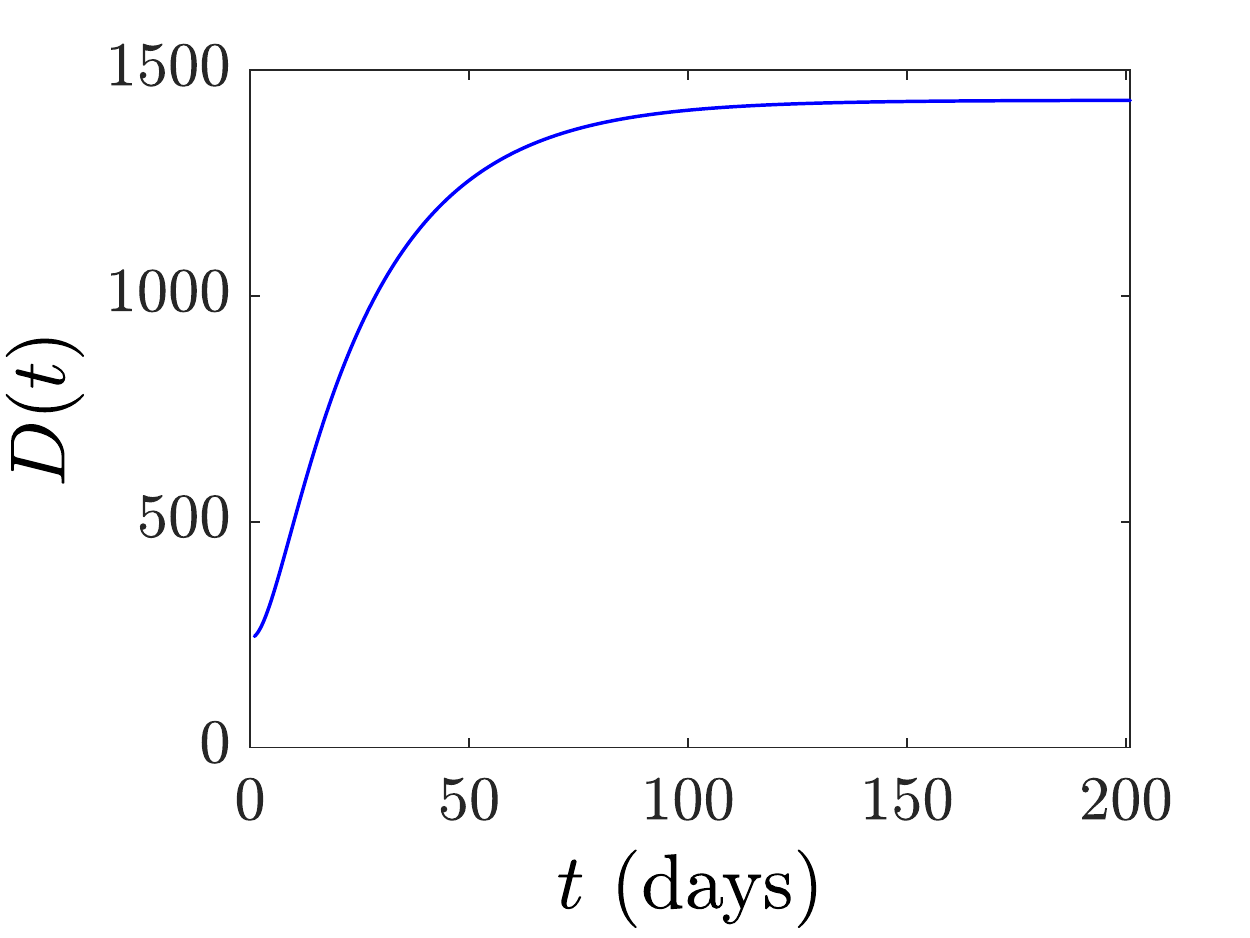} &
 \includegraphics[width=.3\textwidth]{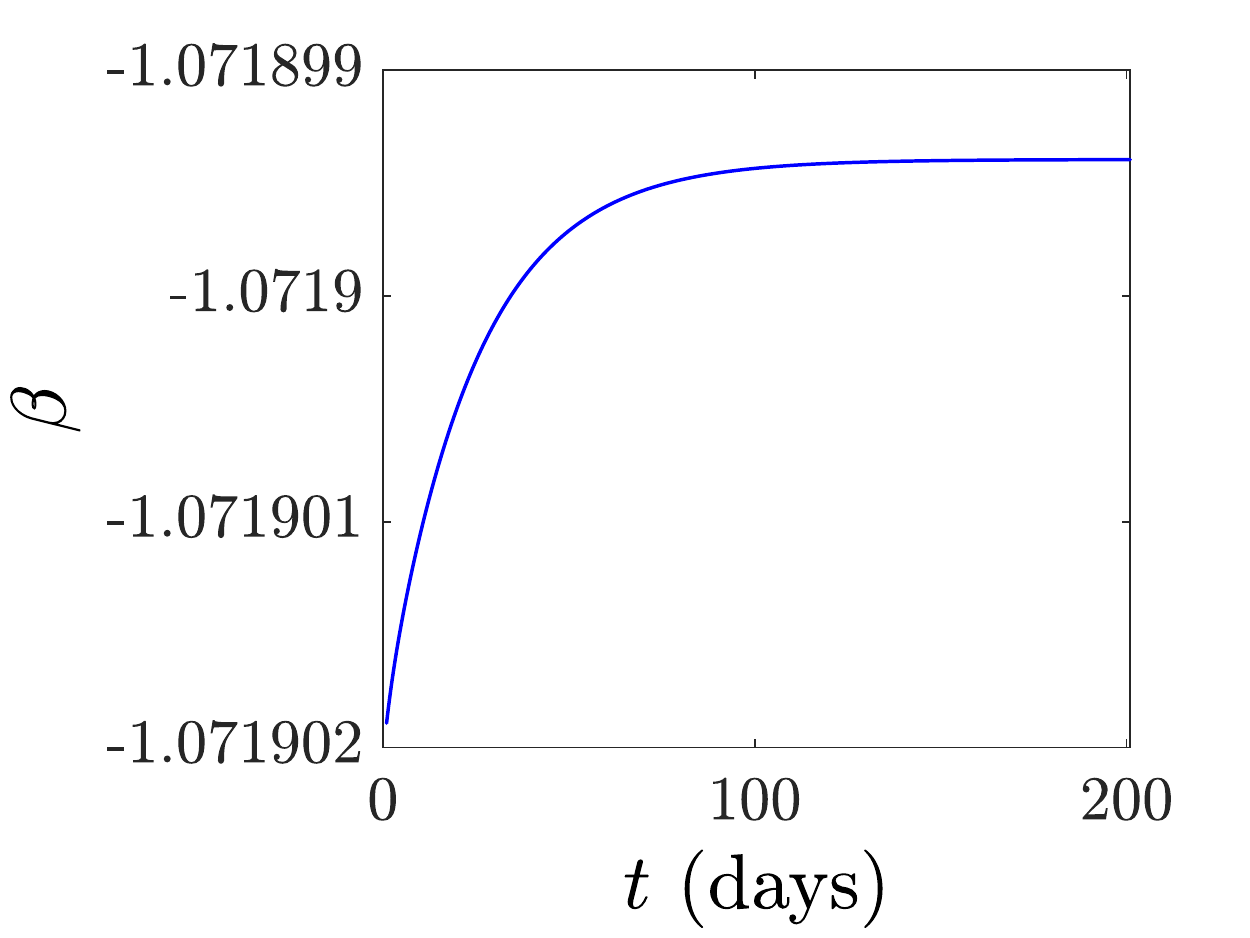}  &  \\
  \includegraphics[width=.3\textwidth]{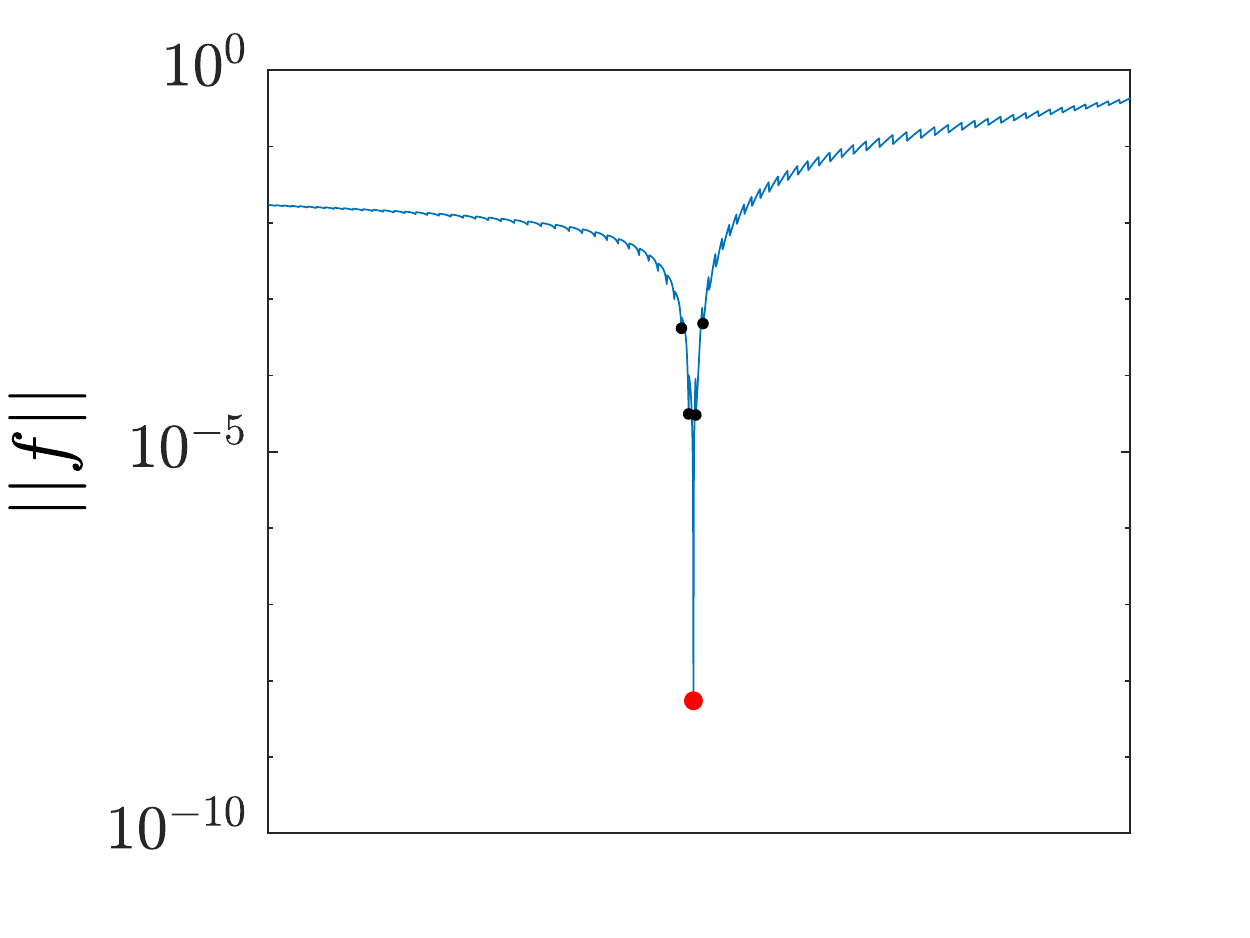} &
  \includegraphics[width=.3\textwidth]{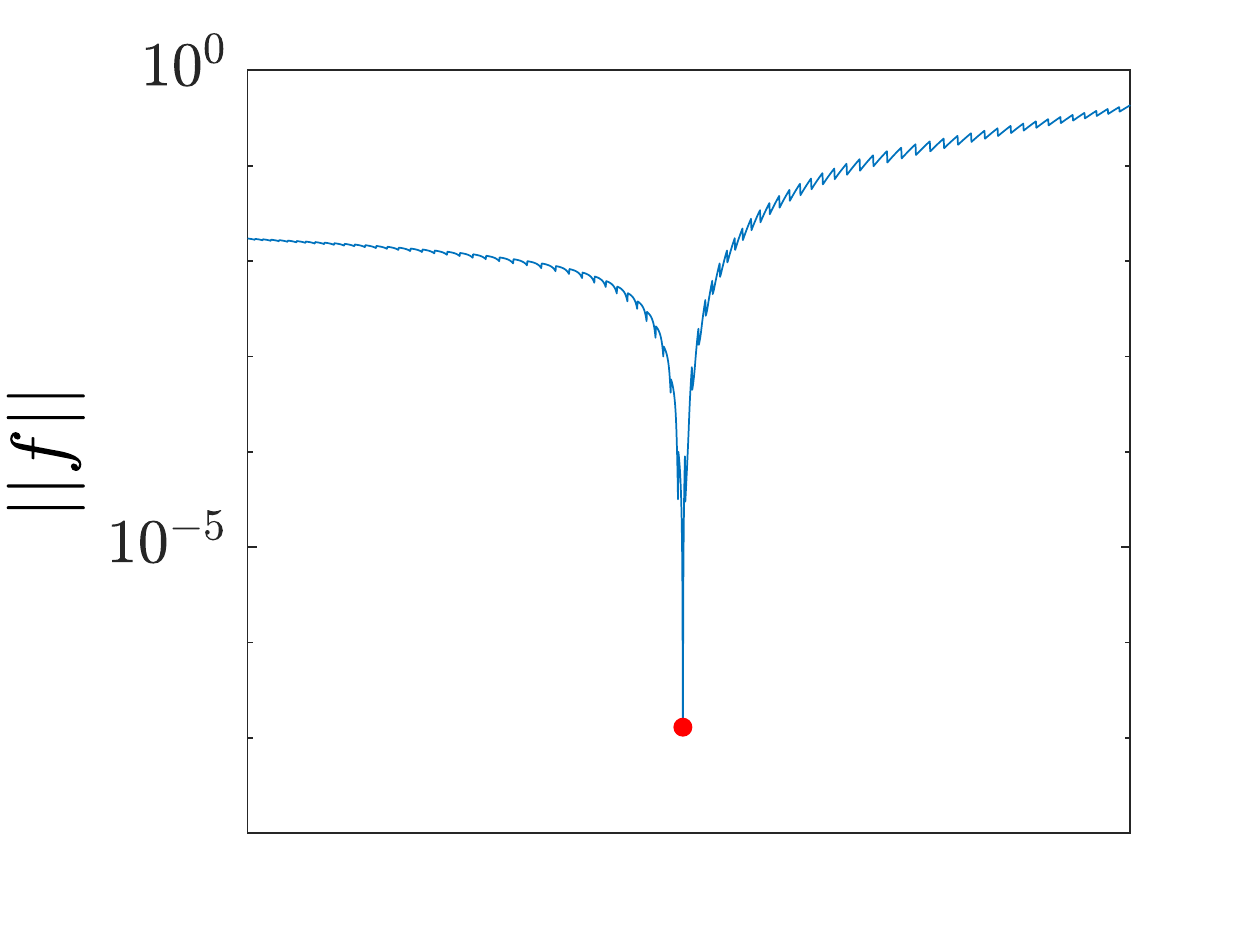}  &
 \includegraphics[width=.3\textwidth]{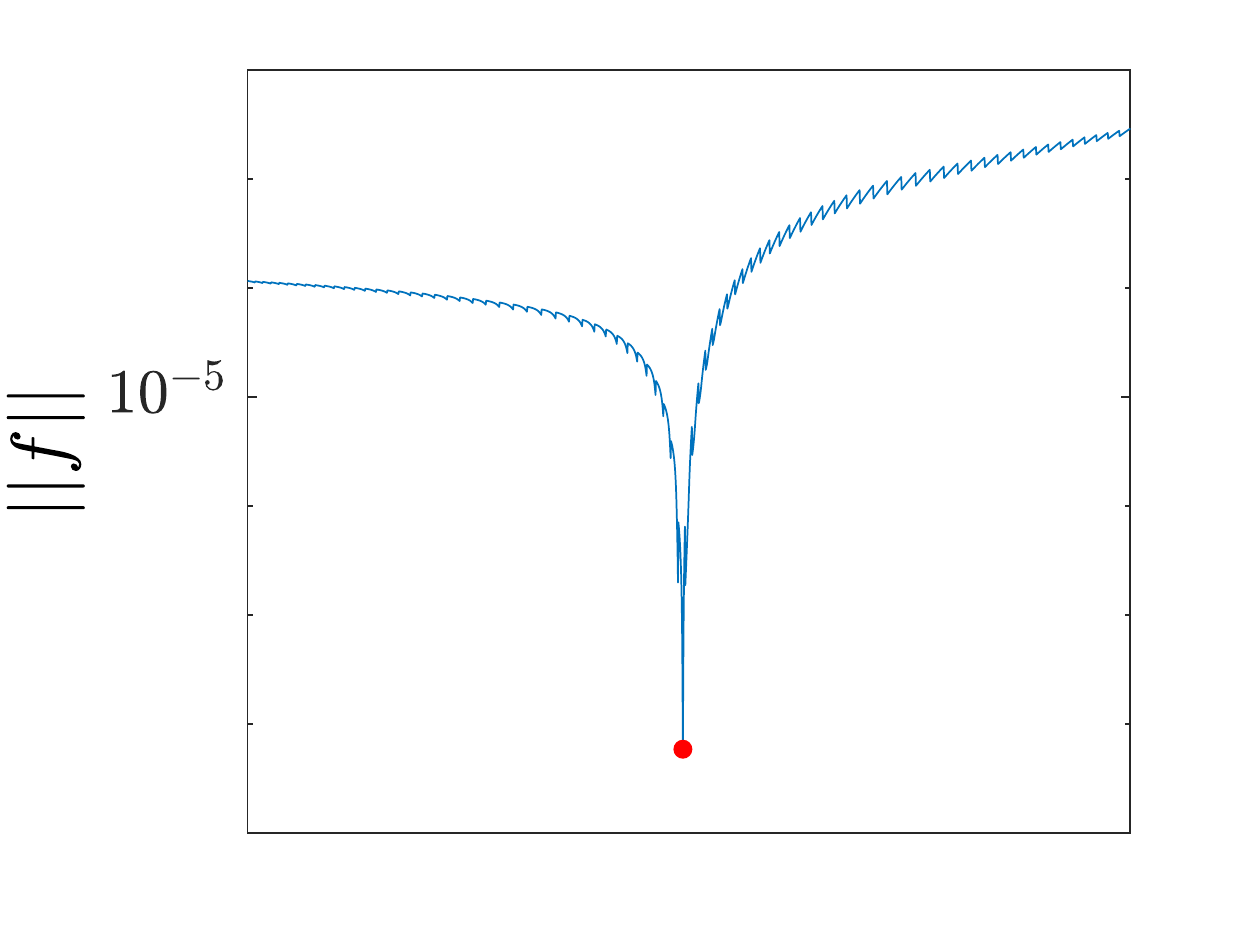}
\end{tabular}
\end{center}
\caption{The top left panel shows the synthetic time series of $D(t)$.
   The top middle panel shows the numerical computation of
    $\beta$ illustrating that indeed as expected this first integral of the motion
    remains invariant over time, up to numerical accuracy.
    In the bottom panel the time series of the left-hand side of Eq.~(\ref{eqn7})
    is evaluated and its $l^2$ norm is calculated. This is done for
     the grid of finely sampled values of $(C_1,C_2)$.
    The red dot indicates the local optimum associated with the known
    parameter values from which the time series was initially constructed. In the left
    panel the rest of the dots correspond to the ones reported in Table~\ref{tab2}
    (the next best fits). The bottom panels correspond to the fitness indicator
    $||f||$ for the exact time series (including its derivatives) on the left, for
    the time series and then its derivatives from finite differences (middle), and from
    a logistic approximation (right). Notice how the identified
    local minimum of the red dot turns out to also be the global minimum
  within the sampling interval considered.}
\label{un_f1}
\end{figure}

      Our first and  benchmark example concerns a synthetic time series,
      inspired by parameter values that are at the proper ballpark for
      describing the real-life data of Portugal's deaths~\cite{porto}:
      That is to say, we prescribed a parameter set of {\it all 9} parameters,
      resulting in the lumped (per our analysis above) coefficients of
      Table~\ref{tab1}. Upon assigning realistic initial conditions, we ran the
      differential equations~(\ref{eqn1})--(\ref{eqn0}) forward, and produced
      the time series of the top left panel of Fig.~\ref{un_f1}. Notice that
      in this case all data $(I(t),A(t),S(t),H(t),R(t),D(t))$ are available (in this procedure, the only
      error involved stems from the approximate nature of our numerical integration); from these quantities
      we  constructed $D^{(1)}$, $D^{(2)}$, $D^{(3)}$ and $D^{(4)}$, again
       within the above (small for our high order numerical scheme) error.
      The first check to which we subjected our method is that of identifying
      $\beta$: if the above time series of $D(t)$ and its derivatives are accurate,
      using Eq.~(\ref{eqn7}) to obtain $\beta(t)$, the corresponding time series
      should be found to be constant. We observed this
      was the case up to the 5th significant digit, as shown in the top middle panel of Fig.~\ref{un_f1}, in line with the high accuracy of our numerical integrator.

      Then, we turned to the inverse problem. Suppose we  are {\it only}
      given at first the time series of $D(t)$ and its derivatives. Can we use
      the time series of $f(t)$ and in particular (practically) the minimization
      of $||f||$ to uniquely identify the reduced, uniquely identifiable
      parameters within the set from which we started? To provide the answer
      for this synthetic example, we initially
      substituted all of these time series to the left-hand side of Eq.~(\ref{eqn7})
      and
      examined whether the minimization of the time-series stemming from the ODE
      provided the expected  local minimum (which was
      a priori known as a result of the  knowledge of all model parameters for this time series).
      Indeed, we found this to be the case, as per the data presented in Table~\ref{tab1},
      under ``exact'' (indicating that this was generated from the exact time series
      --up to the approximations of numerical integration--
      for $D(t)$ and also $D^{(1)}$, $D^{(2)}$, $D^{(3)}$ and $D^{(4)}$).
We then explored a grid of values around this central minimum to decide whether it
is a local or global minimum.
In this grid,
      we included, for {\it each} of the 8 parameters entering
      the minimization, values which corresponded to the minimum (the  minimum will be
      denoted as ``1'' i.e., 1 $\times$ the minimum), $1/2$ denoting $1/2$ of the minimum,
      $1/3$, $2$ and $3$ times the minimum; in this way, we created a tractable ``grid''
      of values within a fairly large range of variation from the ballpark
      values of interest, and posed the following  question:  was the expected from our analysis global minimum the one associated with  the coefficients that we started with
      (the set of 1's  in the language of the multiples considered)
      or not?
      The answer within this table of $5^8=390625$ entries for the present
      example was found to be in the
      affirmative.

In order to obtain a more refined
      sense of whether we really obtained the global minimum, we then went on to explore a finer grid of parameter values.
      To that effect, we decided to explore (and illustrate) the method on the
      basis of a considerably more refined Table but of fewer parameters.
      Through our various explorations of the method, we identified
      as the parameters that were most robust under fitting the ones associated
      with direct rates, namely $R_2$, $R_3$, $r_1$ and $F=a+s$. Also,
      $\alpha$ barely varied due to its being dominated by the contribution
      of $T$ within its definition; similarly with $\beta$.
      In that light, we decided to hereafter focus on the role of $C_1$ and
      $C_2$ that are also, in a sense, some of the most crucial parameters
      of the model in controlling additional infections, the spreading
      of the epidemic (also per Eq.~(\ref{evalue2})) and, hence, ultimately
      the deaths that result from the model. They are also the ones that would
      be  crucially affected under the easing  of lockdown conditions;
      hence, in our view these parameters are the most significant ones to examine in terms of
      the robustness of the result across such variations.
      Among the two ($C_1$ and $C_2$), the crucial
      role of asymptomatics (due to their mobility within the population)
      renders $C_1$ as the typically dominant and significantly larger value.
      In that light, we restricted our considerations to the interval
      $ C_1/10 \leq C_2 \leq C_1/3$ (unless indicated otherwise) in what follows below.
      This was done to avoid combinations of $(C_1,C_2)$ that may lead to
      ‘competing minima’, where $C_2$ is  comparable or larger than $C_1$; this
      would mean that asymptomatics would have a similar or smaller number of contacts
      as the (expected to be self-isolating) sick,
a feature not in line with the epidemiological premise of the model.
Given the above considerations, we decided to consider a refined variation
of $C_1$ over a grid of $(1/5, 1/4.9, 1/4.8,\dots,1/1.1,1,1.1, \dots, 4.8, 4.9, 5) \times$ the original value, and similarly of $C_2$ (within the limits of the above inequality
condition) of the above reported minimum of Table~\ref{tab1}.
The results of this fine sampling variation are shown in the bottom left
panel of Fig.~\ref{un_f1}, as well as in Table~\ref{tab2}. These results confirmed that the
original identified minimum is the global one with the optimal fitness within this
parametric grid.
Table~\ref{tab2} also compares this fitness by means of measuring $||f||$ (which is
what is also represented in Fig.~\ref{un_f1}) not only for this global
minimum (our (1,1) state given by the red spot in this figure and also in the
panels that will follow), but also for
the ‘next best things’,  which are given by black spots in the figure.
The edge of the original minimum as being the global one by some
$4$ orders of magnitude is clearly evident
within this refined grid. This confirmed the benchmark of our expectations for this
essentially exact time series. Indeed, in this case even an extensive search
over arbitrary $C_2$'s (not conforming to the above inequality) would still
not improve the global minimum result.

The next line of attack was to seek a similar optimization approach in this well-curated and smooth time-series
example (again the one of the top left of Fig.~\ref{un_f1}),  but
now assuming that we are {\it not} given both $D(t)$ and all of its
derivatives, but rather only $D(t)$.  Our first approximation
of $D^{(1)}$, $D^{(2)}$, $D^{(3)}$ and $D^{(4)}$ in this smooth case was based
on the numerical method of finite differences. We used this well-known tool to approximate
all derivatives. Then, again, we first  posed the following question: equipped with these time
series, could we find a local minimum of Eq.~(\ref{eqn7}) through minimizing
$||f||$ in the vicinity of the exact result? The answer was again in the affirmative
and is accounted for in the corresponding ``Finite Differences'' column of Table~\ref{tab1}.
It can be seen that despite the approximate nature of the method, we
still effectively captured all relevant (lumped, reduced) coefficients. Subsequently,
we again followed the sampling procedure developed earlier with the fine grid
over the most sensitive and, arguably, most important (towards the easing of the lockdown
measures) parameters, namely $C_1$ and $C_2$. Once again, we found, as seen
in the middle bottom panel of Fig.~\ref{un_f1}, that the original seed,
namely the minimum that was expected, is the one that beats the competition
in terms of its fitness by being the lowest one. However, the caveats
of the approximate nature of the method (when only $D$ is prescribed)
are starting to come into light. In particular, it can be seen that next to the
red spot, corresponding to the $(1,1)$ parametric set, the one
that was expected, there are others that are less clearly discernible than the
previous example in terms
of their fitness. In other words, the edge of our expected optimum was significantly
decreased by the approximate nature of the time series: from an
edge of $4$ orders of magnitude, we went to one of only $2$, as is shown
in first row of Table~\ref{tab3}. As a way to read the relevant tables (also for the sake
of future examples), we note that the fitness of the global minimum in terms of its
$||f||$ is $1.3 \times 10^{-7}$, while the next best thing concerns a
$C_1$ that is $1.1$ times larger and a $C_2$ that is $1.8$ smaller
and yields a $||f||=3.65 \times 10^{-5}$.

This issue comes into sharper focus when we proceeded to the further
approximation of the logistic formula (sigmoid). In the case of a realistic time
series, the data becomes quite noisy and hence the notion of finite differences
introduces a considerable amount of this noise into the minimization of $||f||$.
Hence, the logistic approach is deemed to be a worthwhile alternative, to consider because
of the smoothness of the curve and its derivatives, and the effective transformation
of the left-hand side of $f$ into a well-defined function of $x$ (possibly even
a polynomial upon a suitable expansion of the logarithmic term). However, these
advantages come with some caveats too. As we can see in the synthetic
time series at hand, we can again perform a minimization with the logistic approximation
in the vicinity of the original exact parametric set of this example. We indeed found,
in that case too, a local minimum reflected under the sigmoid column
of Table~\ref{tab1}, which is in good agreement with the previous methods;
notice, however, the slight deviations in the integration constants
due to the approximate nature of the logistic model.
In this progressively more realistic case (since, typically, neither a smooth
time series for $D(t)$ and all its derivatives, nor even a smooth one for $D$ alone
will be practically available), we see that the fitness of our global minimum only
edges that of the next closest minimum by a smaller amount than in the finite
difference case.
However, it is crucial to emphasize that in this example, as well as in the examples to follow, the closest second
minimum, $C_1$ that plays the {\it dominant}
role in determining the fate of dynamics of $D(t)$, is very close to (only $1.1$ times)
that of the Table~\ref{tab1}. Hence, even in a foreseeable case where such a secondary
minimum may eclipse our expected global one, the dynamics of the two cases will be very close to each other.

\begin{figure}[!ht]
\begin{center}
  \begin{tabular}{ccc}
    \includegraphics[width=.3\textwidth]{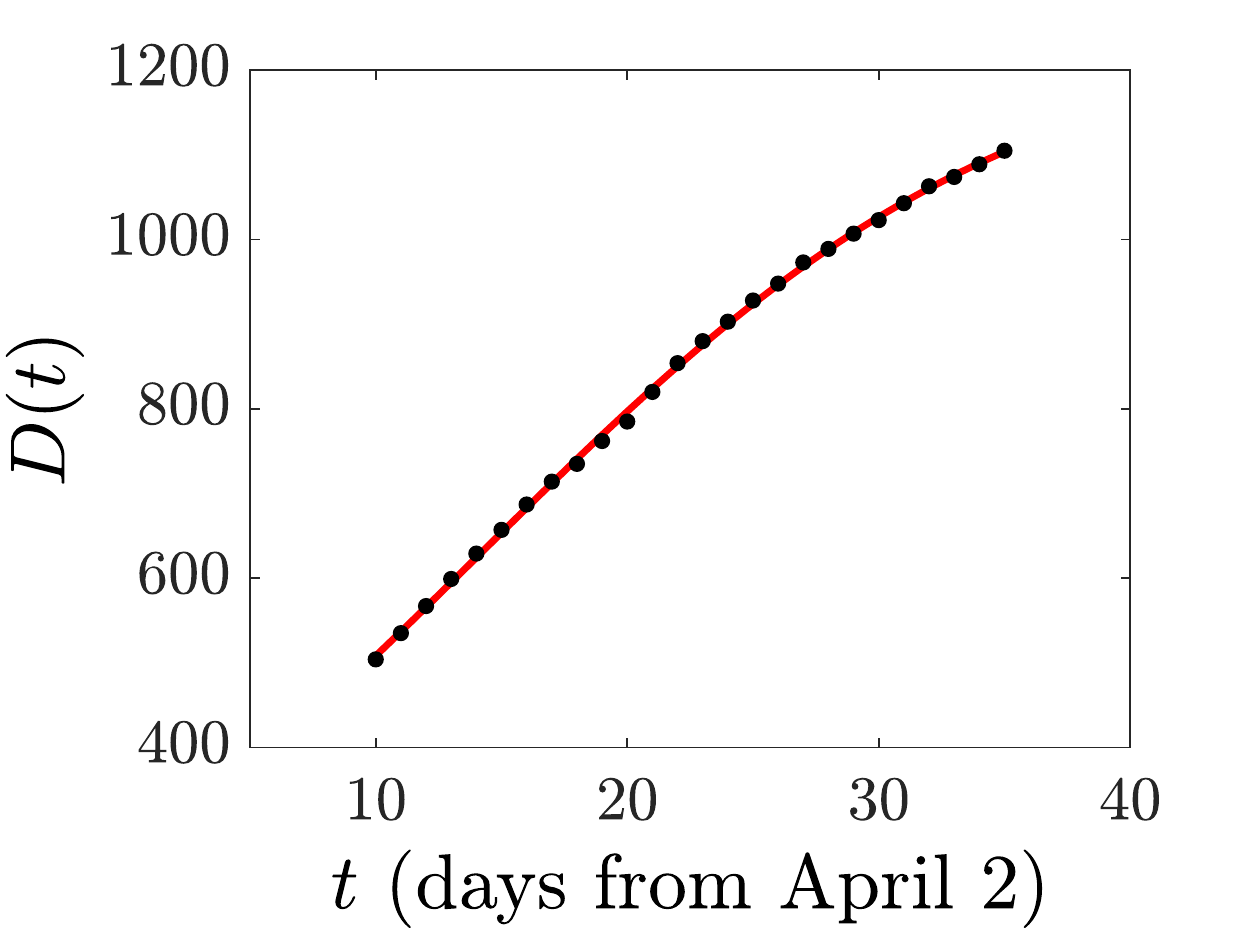} &
\includegraphics[width=.3\textwidth]{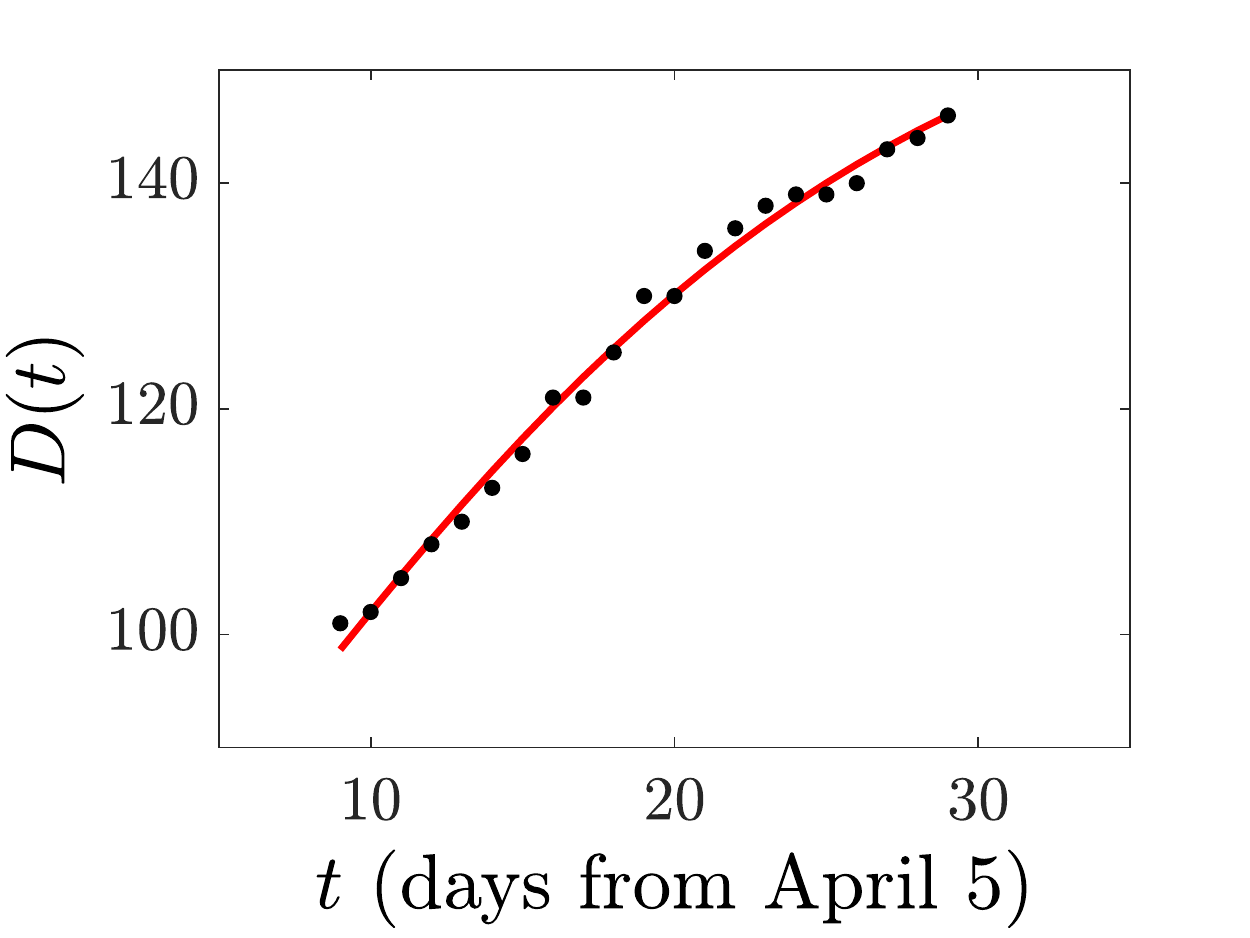} &
\includegraphics[width=.3\textwidth]{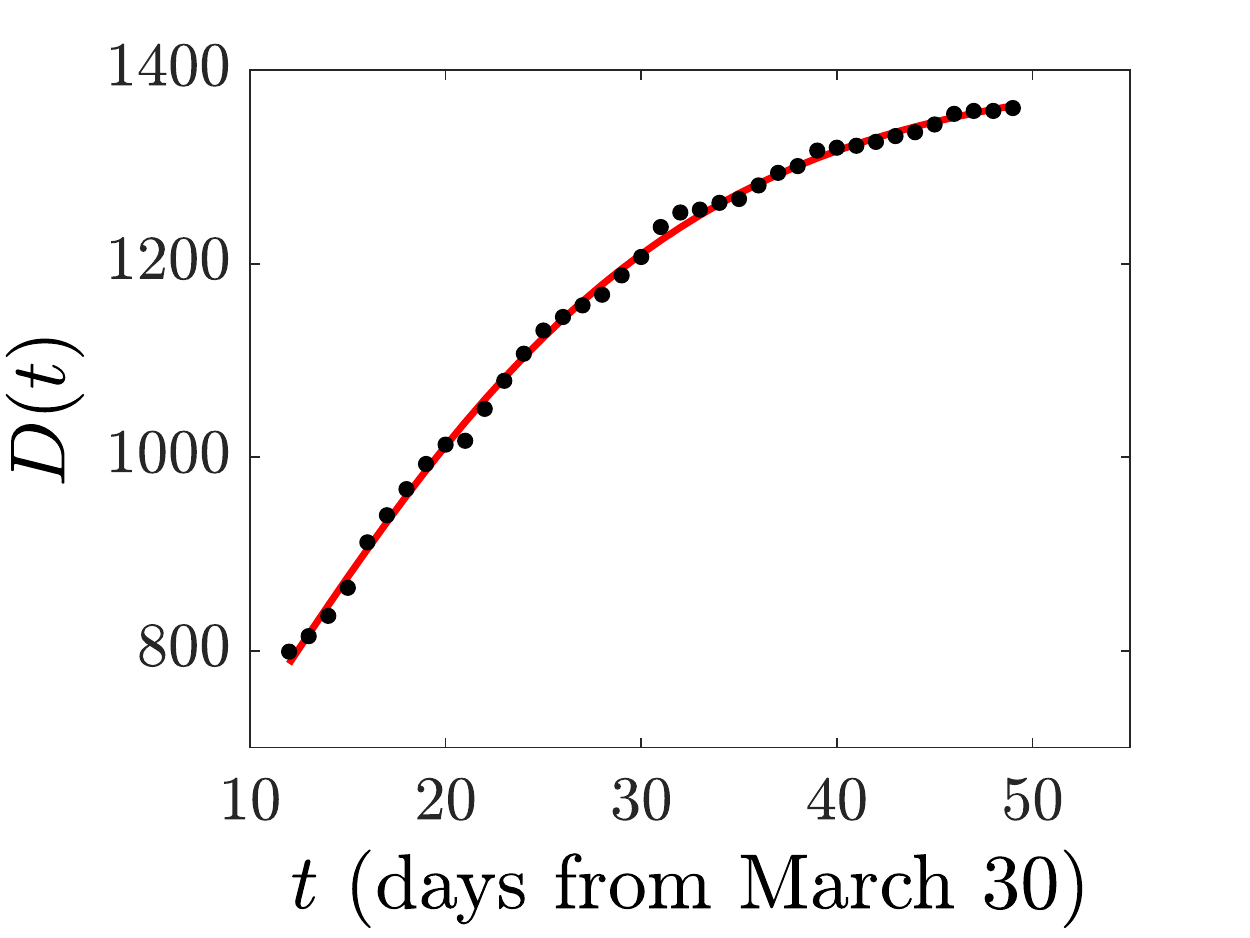}  \\
\includegraphics[width=.3\textwidth]{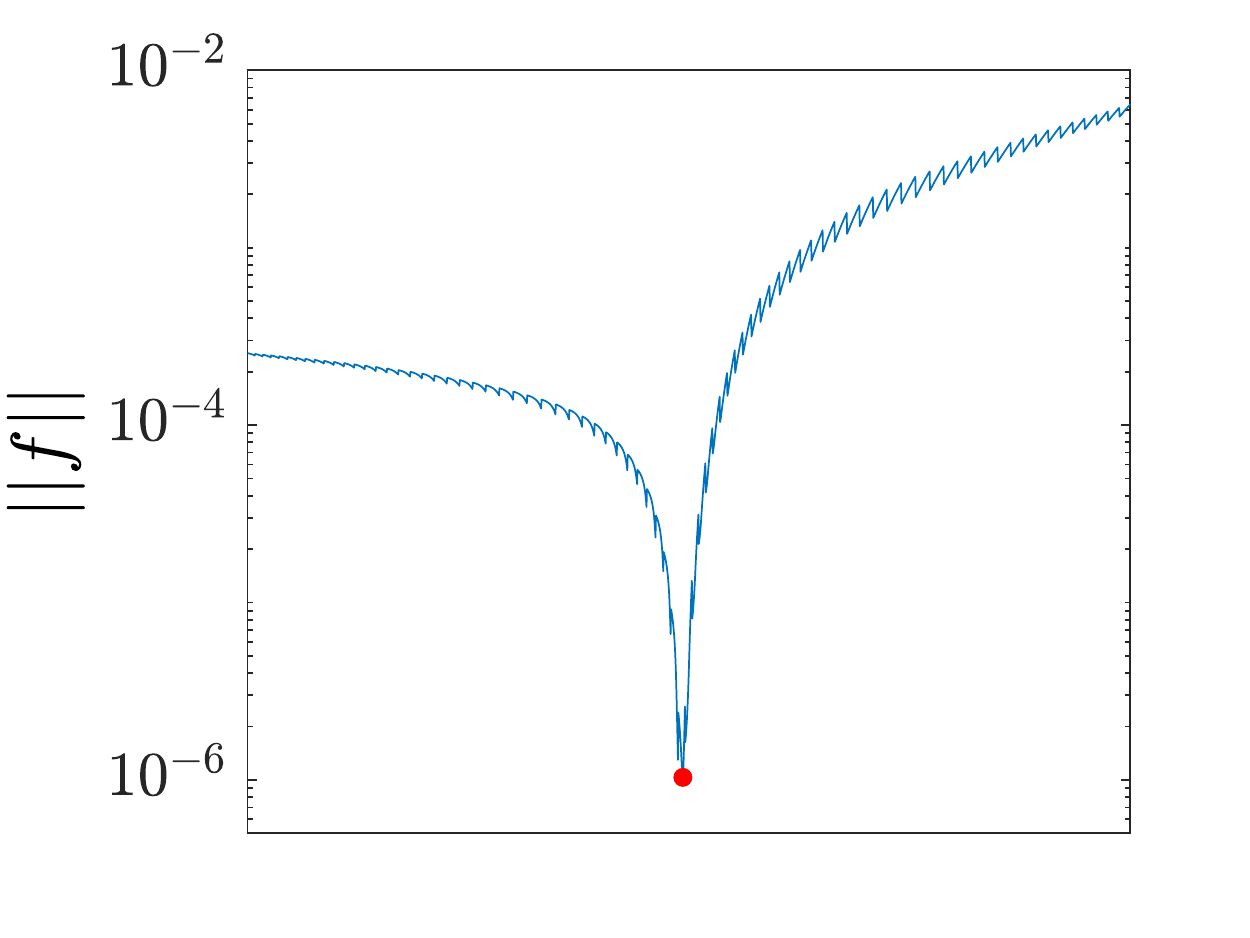} &
  \includegraphics[width=.3\textwidth]{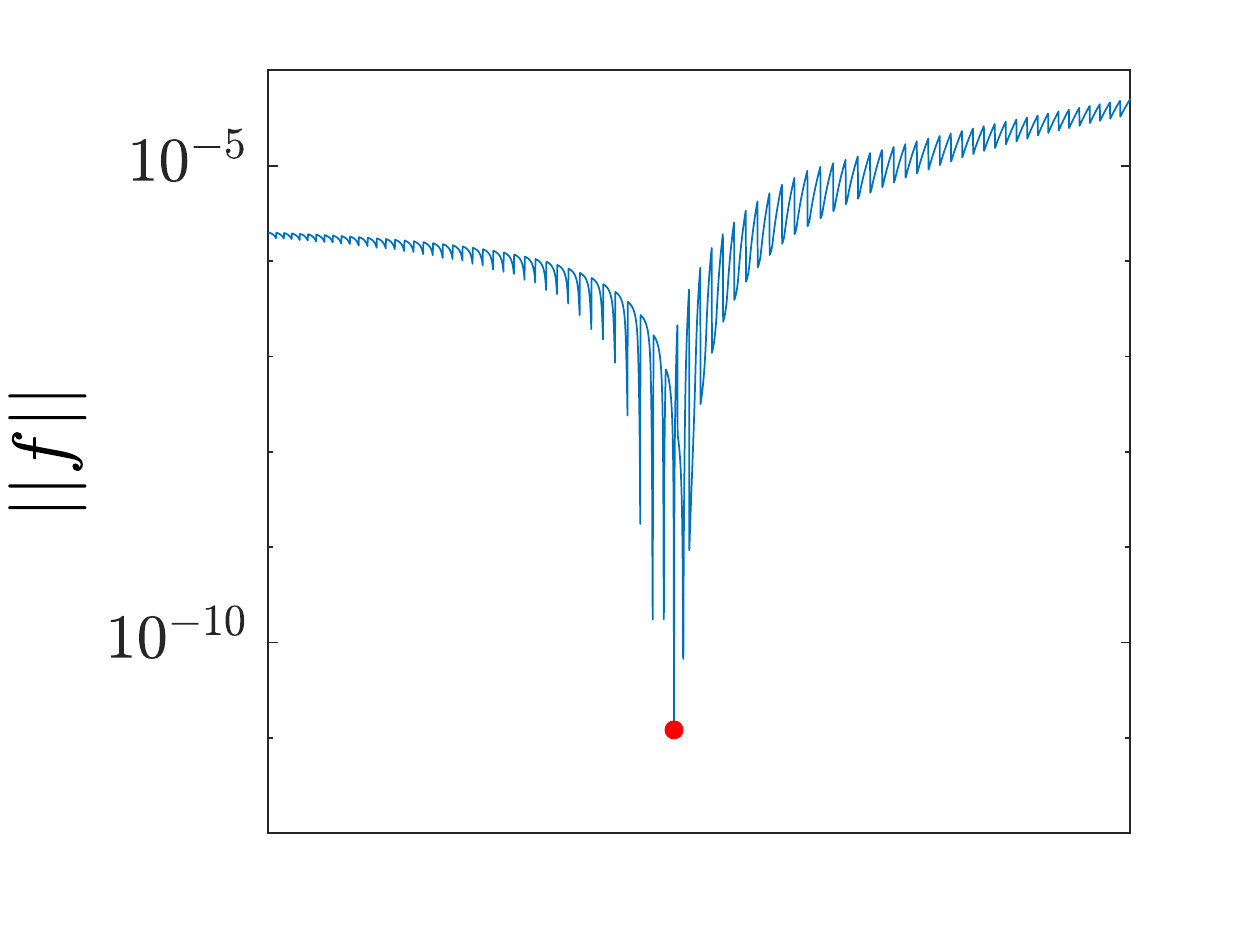} &
  \includegraphics[width=.3\textwidth]{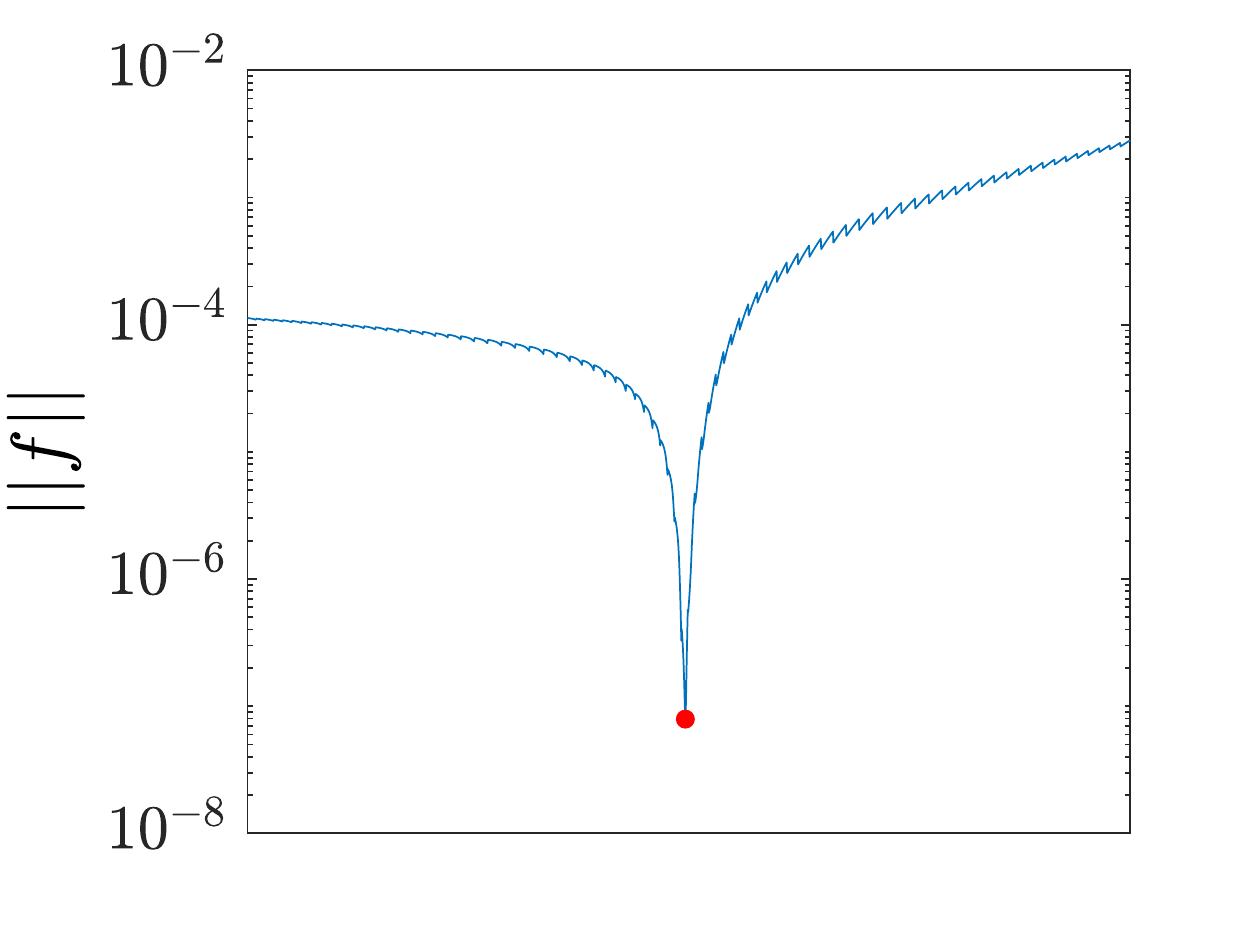}  \\
 \end{tabular}
\end{center}
\caption{The top 3 panels present the data for $D(t)$ for the countries of
  Portugal (left column), Greece (middle column), and the autonomous community of
  Andalusia in Spain (right column). Also shown is a logistic fit of the central
  portion of the associated data aimed at capturing the logistic coefficients $k$ and $D_F$.
  Then in the bottom panel, similarly to the bottom panels of Fig.~\ref{un_f1},
  but now only for the logistic approximation of the top panel, the optimum within
  a refined grid of values of the $C_1$ and $C_2$ parameters is traced; the
  corresponding seed at the center of our consideration is represented by
  the parametric
  set found by constrained minimization and shown by a red dot
  in each case.}
\label{un_f2}
\end{figure}

As may be anticipated, similar issues arose
when we attempted to utilize the logistic
approach to real data for the 3 examples outlined in the main text: the
countries of Portugal (left panels of Fig.~\ref{un_f2}), Greece
(middle panels of Fig.~\ref{un_f2}) and the autonomous community
of Andalusia in Spain (right panels of Fig.~\ref{un_f2}). In each case,
we attempted to fit a portion of the data, the one past the inflection point
in order to capture as best as possible the asymptotic number of deaths
$D_F$ which is a crucial aspect of the sigmoid (together with the
growth coefficient $k$). In our experience, $k$ is usually the most
robust feature of the associated sigmoids; $D_F$ is the trickiest
one to obtain, especially with time series such as the ones of Fig.~\ref{un_f2}
which have not clearly asymptoted to a concrete value. Typically, we have
found the sigmoid to under-predict $D_F$. We have identified this
as one of the significant deficiencies of the sigmoid which, in turn, calls for improvement
of the method, e.g., along the lines of~\cite{FDK1,FDK2}.
The associated fits to sigmoids for the 3 examples are shown in the top
panels of Fig.~\ref{un_f2}, while the associated sigmoid parameters are shown
in Table~\ref{tab3}.
Although in these examples the use of the sigmoid was
sufficient, developing the relevant extension along the lines of e.g. the rational
approximation~\cite{FDK1,FDK2} may be a natural vein for future studies
and may indeed enhance the robustness of the method proposed herein
(especially in the presence of noisier or more sparse time series).

The bottom panels of the figure illustrate further what may happen in the
case of the sigmoid. Notice that in the case of these real data for Portugal (left),
Greece (middle) and Andalusia (right), we have
no a priori knowledge of what the optimum fit parameters are. However,
as a comparison measure, we utilized the parameter values found by a constrained minimization
along the lines discussed in~\cite{FCK} (see also~\cite{preprint}).
Admittedly, this minimization using Matlab's routine  \texttt{fmincon}
(for minimizing the distance between true and observed time series stemming
from integrating the differential equations)
does not offer any guarantee of a global minimum and indeed
may yield a local minimum. Nevertheless, it provides a useful benchmark for
the performance of the minimization of $||f||$ within Eq.~(\ref{eqn7}) by using
the logistic approximation. Subsequently, based upon the logistic fit
(most notably its $k$ and $D_F$), we attempted to identify a local minimum
through the logistic method in the vicinity of the one obtained by constrained minimization.
Indeed, such a minimum was found in each of the 3 cases;
and its parametric values are illustrated in
Table~\ref{tab4}. Then, we attempted again a refined grid of $(C_1,C_2)$  conforming to
the same constraint as before, except for the case of Greece: there,
given the result of the constrained minimization for $C_2 \approx C_1/3$, we
extended our search to a wider parametric interval of $0.1 C_1 \leq C_2 \leq C_1$
for completeness. Indeed, we checked that our results for this case did
not change even for a wider parametric search.

In all the explored cases, we found the local minimum (the one obtained
both by constrained minimization and independently by the logistic approach and
presented in Table~\ref{tab4}) to be the global minimum within the interval of our
search. Nevertheless, once again, it is clear that other combinations of
$C_1$ and $C_2$ may yield fitnesses close to that minimum, since
the differences under the logistic approximation were always less
than one order of magnitude. Naturally, this raises  concerns regarding
the identification of the global minimum. However, we note the following: (i) In this case, we have checked (something that is possible more generally) the result in the
interval of interest under two separate methods. (ii) More importantly, as noted earlier, the secondary minima identified
involve similar results as regards $C_1$ and hence are expected to have
similar dynamical implications. (iii) More refined variants of the
method (e.g., using the rational or
birational models~\cite{FDK1,FDK2}) should be able to address the needs of more
complex/less informative cases than the ones discussed here. In that vein, we believe
that we have exposed both the advantages and the caveats of the
proof-of-principle logistic approach and how it can be improved, as needed.

\begin{table}
\centering
\begin{tabular}{ccccc}
Series & $\overline{C_1}$ & $\overline{C_2}$ & $||f||$ & $||f||$ (global) \\
\hline
Synthetic (Finite differences) & $1.1$ & $1/1.8$ & $3.65\times10^{-5}$ & $1.30\times10^{-7}$\\
Synthetic (Sigmoid) & $1.1$ & $1/1.9$ & $1.89\times10^{-7}$ & $5.91\times10^{-9}$\\
Real series (Portugal) & $1/1.1$ & $1.4$ & $1.30\times10^{-6}$ & $1.03\times10^{-6}$\\
Real series (Greece) & $1.1$ & $1/1.6$ & $6.74\times10^{-11}$ & $1.21\times10^{-11}$ \\
Real series (Andalusia) & $1/1.1$ & $1.5$ & $3.28\times10^{-7}$ & $7.91\times10^{-8}$ \\
\hline\\
\end{tabular}
\caption{Similar to Tab.~\ref{tab2} but now for each of the remaining 5
  examples (corresponding to the 5 rows). Here the fitness of the global
  optimum is shown in the 5th and last column, while the next best thing
  (in multiples of the optimal $(C_1,C_2)$) and its associated fitness
are shown in columns $(2, 3)$ and $4$, respectively.}
\label{tab3}
\end{table}

\begin{table}
\centering
\begin{tabular}{cccc}
Parameter & Portugal & Greece & Andalusia\\
\hline
$D_f$ & $1264.0$ & $162.93$ & $1404.4$ \\
$k$ & $0.0931$ & $0.0863$ & $0.0881$ \\
$P$ & $1.4874$ & $0.6516$ & $0.7855$ \\
$C_1$ & $2.3687\times10^{-5}$ & $1.5486\times10^{-5}$ & $9.4940\times10^{-6}$ \\
$C_2$ & $5.0879\times10^{-6}$ & $6.4947\times10^{-6}$ & $1.8864\times10^{-6}$ \\
$R_2$ & $0.2451$ & $0.2467$ & $0.2964$ \\
$R_3$ & $0.1309$ & $0.1175$ & $0.1636$ \\
$F$ & $0.3815$ & $0.6100$ & $0.6798$ \\
$\alpha$ & $1390.8$ & $1474.7$ & $5436.7$ \\
$r_1$ & $0.1482$ & $0.1464$ & $0.1426$\\
$\beta$ & $-1.0723$ & $-1.0679$ & $-1.2260$\\
\hline\\
\end{tabular}
\caption{Similarly to Table 1 (for Fig.~\ref{un_f1}), but now for the 3
  sigmoid-based country/autonomous province examples of Fig.~\ref{un_f2},
  the lumped parameter combinations of the model are given, together with the
coefficients of optimal sigmoid fit performed within the relevant figure.}
\label{tab4}
\end{table}

\section{Conclusions and Future Recommendations}

We conclude with several remarks, summarizing our work and offering
a number of future recommendations and interesting research directions:

\begin{enumerate}
\item We have introduced a rigorous and computationally tractable methodology for computing the impact on the number of deaths of increasing the parameter $c_1$ in our SIR-type model.
  This parameter is proportional to the number of contacts between asymptomatic individuals infected with
SARS-CoV-2 and those
susceptible. Remarkably, the only data needed for our algorithm to be
implemented is the cumulative number of deaths during the lockdown period.
  Our considerations have been based on an extended version of an SIR model
  (incorporating asymptomatically infected, hospitalized individuals, as well
  as deaths). Using an inverse problem perspective, we have formulated a single
  ODE for the fatalities, which we consider as the ``ground truth'' within the
  available data. Via an optimization scheme, we have utilized this $D(t)$
  (and its derivatives, obtained, e.g., via a logistic fitting to the data) to
  obtain the uniquely identifiable parametric combinations within the model.

\item In addition to providing synthetic examples to illustrate the method,
  we have also utilized real data in order to showcase the practical implementation
  of the proposed approach.
  The application of our algorithm towards the study of the
  pandemic spread to Greece, Portugal and Andalusia shows  that upon easing of the lockdown  measures by increasing the value of $c_1$ (and hence
  effectively the number of contacts) by a multiple larger than 2,  the impact on the number of deaths will be devastating. However, an increase by $\zeta=2$
  has a more modest effect, especially in the case of Greece which is found to have
  a low transmission rate.

\item Importantly, a similar analysis can be applied to sub-populations. Let us consider two such subpopulations: one consisting of individuals below the age of 40, that will be referred to as young, and one consisting of individuals above 40, that will be denoted as older. If there exist reliable deaths data for these two subpopulations, then it is possible to compute separately the effect of changing the interactions among the young, as well as among the older with the older and the older with the young.
    We did, for example, look at these two subpopulations in the case of
    Greece~\cite{FCK}. In this case, we supplemented the data of deaths with data on the cumulative number of reported infected, because the data on deaths, especially for the young, were sparse. These results provided hope (and a potential future
    recommendation, namely) that it may be possible to accelerate the easing of the lockdown for the young: the increase of interactions among the young (as opposed to the interactions involving the older) led to only a modest increase in the number of deaths. However, it should be noted that, as discussed above, our conclusions for single populations are based on rigorous mathematical results and reliable data; since this is less transparently so the case for the data (e.g., for
    the reported infected) regarding the two sub-populations of Greece, the latter results can only be considered as suggestive. From a mathematical perspective,
    completing and practically applying
    the rigorous analysis in the same spirit as above for two or
    more populations is certainly a challenging problem for future studies.

  \item Our algorithm can be used in any country with reliable death data during the lockdown period; if death data also exist for sub-populations, the multicomponent   version of our algorithm should also be applicable. It would constitute
    an interesting future recommendation to apply our model widely to a number
    of other case examples, in order to identify the parametric differences
    between countries with high and ones with lower fatality case numbers.

\item A central mathematical finding of our analysis concerns the
  rigorous illustration of the fact that the solution of the inverse problem
  (based on death data) does not yield in a unique way all the model parameters, thus it is not directly possible to compute the effect of the easing of lockdown on the number of hospitalizations and on the total number of infected individuals.
  However, we can compute the arguably most important feature of the epidemic,
  upon release of lockdown, namely the number of anticipated future deaths on the
  basis of the model.

\item At the moment, we cannot provide a specific relation between concrete protective measures (such as hygiene conditions, partial social distancing, and perhaps
  especially wearing a
  mask) and the value of $\zeta$. Such a relation exists, as $c_1$ does not
  solely reflect the number of contacts but also the probability of infection
  given a contact which is proportional to the viral load (i.e.,
  the viral concentration in the respiratory-tract fluid) of expelled respiratory
  droplets~\cite{Editorialyd2020}. Clearly, the above protective measures reduce the
  viral load and therefore $c_1$.
  Hopefully, such a relation can be established by designing specific experiments
  in the near future. This is a future recommendation that would be especially
  valuable towards assessing the impact of models within the broad class
  considered herein.

\item Our analysis makes the crucial assumption that the  basic
characteristics of the virus in the post-lockdown stage are identical
with those during the lockdown. If, for example, a mutation takes place
which makes the virus weaker, then the effect of easing the lockdown
measures will not be as dramatic as predicted in our work.

\end{enumerate}

{\bf Acknolwegdments.} ASF acknowledges support from EPSRC in the form of a a Senior
Fellowship. PGK and JCM greatly appreciate numerous helpful discussions with
Y. Drossinos, Z. Rapti, and G.A. Kevrekidis.
This material is based upon work supported by the US National Science
Foundation under Grants No. PHY-1602994 and DMS-1809074
(PGK). PGK also acknowledges support from the Leverhulme Trust via a
Visiting Fellowship and thanks the Mathematical Institute of the University
of Oxford for its hospitality during part of this work. The
work presented here is part of a project that we begun jointly with
Nikos Dikaios, with whom we hope to present elsewhere other aspects of
this project.

\end{document}